\pdfoutput=1
\documentclass[aps,prc,twocolumn,floatfix,superscriptaddress,nofootinbib,preprintnumbers,longbibliography,amsmath,amsthm,amssymb]{revtex4-2}
\usepackage{graphicx}
\usepackage{amsfonts}
\usepackage{color}
\usepackage[colorlinks=true,linkcolor=blue,citecolor=blue]{hyperref}
\usepackage{dcolumn}
\usepackage{bm}
\usepackage{braket}
\usepackage[normalem]{ulem}  

\begin{document}
\allowdisplaybreaks[1]
\title{Triaxial-shape dynamics in the low-lying excited $0^+$ state: Role of the collective mass}
\author{Kouhei Washiyama}
\email[E-mail: ]{washiyama@nucl.ph.tsukuba.ac.jp }
\affiliation{Center for Computational Sciences, University of Tsukuba, Tsukuba, Ibaraki 305-8577, Japan}
\affiliation{Research Center for Superheavy Elements, Kyushu University, Fukuoka 819-0395, Japan}
\author{Kenichi Yoshida}
\email[E-mail: ]{kyoshida@rcnp.osaka-u.ac.jp}
\affiliation{Research Center for Nuclear Physics, Osaka University, 
Ibaraki, Osaka 567-0047 Japan}
\affiliation{RIKEN Nishina Center for Accelerator-Based Science, Wako, Saitama 351-0198, Japan}
\date{\today}
\begin{abstract}
\begin{description}
\item[Background]
Non-yrast states in neutron-rich nuclei are being investigated experimentally. 
These states reveal various aspects and details of the nuclear structure, 
such as the fluctuation around the axially symmetric shape. 
\item[Purpose]
The beyond-mean-field effects in neutron-rich nuclei with $N \simeq 28$ are investigated.
We focus on the role of collective mass in triaxial-shape dynamics.
\item[Method] 
We employ the five-dimensional quadrupole collective Hamiltonian method with the potential obtained in a constrained Hartree--Fock--Bogoliubov approach with a Skyrme energy-density functional and the collective-mass functions obtained by the cranking approximation. The method includes triaxial deformations.
\item[Results]
We find that 
$^{42}$Mg, $^{40}$Si, $^{44}$S, and $^{46}$S show $\gamma$-soft: A flat behavior in the potential energy surface along the triaxial deformation.
Their low-lying spectra show a strong nucleus dependence,
while those obtained with a collective mass assumed as constant 
are similar to each other.
The energy ratio $E(0_2^+)/E(2_1^+)$ and the $B(E2)$ ratio $B(E2;0_2^+\to 2_1^+)/B(E2;2_1^+\to 0_1^+)$ show a unique property of the $0_2^+$ state,
while the energy and $B(E2)$ ratios in neutron-deficient $\gamma$-soft nuclei with $N=78$ do not depend on nucleus so much. 
\item[Conclusions]
Low-lying spectra are determined by not only the potential energy but also the collective mass. 
We clarify the important role of the collective mass in low-energy dynamics in the neutron-rich $N\sim28$ nuclei.

\end{description}
\end{abstract}

\maketitle

\section{Introduction}

Atomic nuclei exhibit various shapes according to the neutron number, 
the proton number, and the excitation energy.
Since the ingredients are finite and their orbital motion is described by quantum mechanics, 
it is essential to consider the fluctuation in shape.
The five-dimensional quadrupole collective Hamiltonian (5DCH) \cite{libert99,prochniak04,niksic09,delaroche10,Matsuyanagi:2016gyp}
as a function of the quadrupole deformation parameters $\beta$ and $\gamma$
has often been employed in describing low-energy states.
The parameters in the model are 
the potential energy and the collective masses.

The shape dynamics is governed by the potential energy at first glance.
In nuclei near the magic numbers, the potential energy surface (PES) 
shows the existence of the local minimum at the spherical configuration.
As the system moves away from the magic numbers, it becomes deformed, 
where one mostly has the local minimum
at the prolately-deformed configuration.
Some nuclei show a soft PES against the triaxial deformation.
Consequently, the low-lying $2^+_2$ state 
or the so-called $\gamma$ vibration shows up.

The vibration in the triaxial deformation is not always harmonic.
An ideal situation where the PES against the $\gamma$ direction is flat 
is investigated by the Wilets--Jean model~\cite{wilets56}. 
In this model, the mass parameters are assumed to be constant.
A characteristic feature of the low-lying states is that the $2_2^+$ state degenerates with the $4^+_1$ state and is lower than the $0_2^+$ state 
among two-phonon states in a spherical harmonic-oscillator potential 
and that the $0_2^+$ is degenerate with the $3_1^+$, $4_2^+$, and $6_1^+$ states.
In addition to the flatness in the $\gamma$ direction, another ideal situation where the PES against the $\beta$ direction is flat is investigated in terms of the E(5) critical point symmetry~\cite{iachello00,casten00,casten06,bonatsos04}.
The PES in the $\beta$ direction is described by an infinite square-well potential.
The degeneracy of the $2_2^+$ and $4_1^+$ is the same as in the Wilets--Jean model, while the $0_2^+$ state is not necessarily degenerate with the $3_1^+$, $4_2^+$, and $6_1^+$ states because of the fluctuation in the $\beta$ direction.

Neutron-rich nuclei around $N=28$ have attracted interest both experimentally~\cite{sohler02,longfellow21,force10,santiago-gonzalez11,gaudefroy09,mijatovic18,longfellow20,campbell06,bastin07,takeuchi12,crawford19}
and theoretically~\cite{rodriguez-guzman02,rodriguez11,egido16,yao11,kimura13,suzuki21,suzuki22,utsuno12,utsuno14,yoshida22,li11,Yoshida:2009jn}.
The authors in Refs.~\cite{rodriguez11,li11,suzuki22} found that 
the breaking of $N=28$ magicity in $^{44}$S
is due to a flat potential, which brings about a wide configuration mixing in the $\beta$--$\gamma$ deformation space.
In a neighboring nucleus $^{43}$S,
the coexistence of prolately-, triaxially-, and oblately-deformed states 
was predicted due to the breaking of $N=28$ magicity in Ref.~\cite{kimura13}.
In $N=26$ and $30$ isotones, it was found that including the triaxial degree of freedom lowers the energy of the $2_2^+$ state~\cite{suzuki21}.
The shape coexistence of prolate and oblate configurations 
and that of oblate and spherical configurations 
were predicted in $^{40}$Mg and $^{42}$Si, respectively~\cite{suzuki22}.
The $\gamma$ vibration was predicted to appear for the 
prolate configuration in $^{40}$Mg~\cite{Yoshida:2009jn}.
Those studies have shown that the triaxial deformation plays an important role in low-energy dynamics in $N \sim 28$ nuclei. 

Not only the PES but also the mass parameters may play a role in describing the low-energy dynamics in a collective Hamiltonian approach.
The mass parameters, as well as the potential energy, depend on the deformations.
The terrain of the PES 
and the deformation-dependence of mass parameters 
are sensitively determined by the shell effect. 
Thus by changing the neutron/proton number involving neutron-rich and 
neutron-deficient nuclei, 
one can investigate the role of the mass parameters in low-energy dynamics. 
Therefore, we study in the present work the low-energy dynamics governed by triaxial deformation in neutron-rich nuclei around $N=28$, putting 
a focus on the role of the mass parameters in the collective Hamiltonian.
 
The paper is organized as follows.
In Sec.~\ref{sec:method}, we briefly explain the 5DCH method. In Sec.~\ref{sec:result}, we show the results and discuss the roles of the mass parameters. Section~\ref{sec:summary} summarizes the paper.

\section{Method}\label{sec:method}

We briefly explain the five-dimensional quadrupole collective Hamiltonian method. For details, we refer to Refs.~\cite{libert99,prochniak04,niksic09,delaroche10,Matsuyanagi:2016gyp}.

The collective Hamiltonian reads 
\begin{equation}\label{eq:CH}
    H = T_{\text{vib}} +T_{\text{rot}} + V(\beta,\gamma),
\end{equation}
with the vibrational and rotational kinetic energies,  
\begin{align}
    T_{\text{vib}} &= \frac{1}{2}D_{\beta\beta}(\beta,\gamma) \dot{\beta}^2+ D_{\beta\gamma}(\beta,\gamma) \dot{\beta}\dot{\gamma} + \frac{1}{2} D_{\gamma\gamma}(\beta,\gamma)\dot{\gamma}^2, \label{eq:kinetic_vib} \\
    T_{\text{rot}} & = \frac{1}{2}\sum_{k=1}^3 \mathcal{J}_k(\beta,\gamma)\omega_k^2. \label{eq:kinetic_rot}
\end{align}
The functions $D_{\beta\beta}(\beta,\gamma)$, $D_{\beta\gamma}(\beta,\gamma)$, and $D_{\gamma\gamma}(\beta,\gamma)$ denote the vibrational masses and 
$\mathcal{J}_k(\beta,\gamma)=4\beta^2 D_k(\beta,\gamma) \sin^2(\gamma-2\pi k/3)$ and $\omega_k$ denote the rotational moments of inertia and rotational angular velocities in the body-fixed frame of a nucleus.
After quantizing the Hamiltonian, we obtain 
the collective Schr\"odinger equation as
\begin{align}\label{eq:collective_schrodinger}
[\hat{T}_{\text{vib}}+\hat{T}_{\text{rot}} + V(\beta,\gamma)]\Psi_{\alpha IM}(\beta,\gamma,\Omega)=E_{\alpha I}\Psi_{\alpha IM}(\beta,\gamma,\Omega),
\end{align}
where $E_{\alpha I}$ and $\Psi_{\alpha IM}(\beta,\gamma,\Omega)$ are the excitation energies and the collective wave functions with the total angular momentum $I$, its $z$ component $M$ in the laboratory frame, and $\alpha$ distinguishing the states with the same $I$ and $M$. 
The collective wave functions are
functions of $\beta$, $\gamma$, and the three Euler angles $\Omega$ and are written as
\begin{align}\label{eq:collective_wavefunction}
    \Psi_{\alpha IM}(\beta,\gamma,\Omega)=\sum_{K=\text{even}}^I \Phi_{\alpha IK}(\beta,\gamma)\braket{\Omega|IMK},
\end{align}
with $\braket{\Omega|IMK}$ being linear combinations of the Wigner rotational wave functions and $K$ being the $z$ component of $I$ in the body-fixed frame.
The vibrational wave functions $\Phi_{\alpha IK}(\beta,\gamma)$ are normalized as
\begin{multline}\label{eq:normalization}
    \int_0^\infty d\beta \int_0^{\pi/3} d\gamma |G(\beta,\gamma)|^{1/2} \\
    \times \sum_{K=\text{even}}^I
    \Phi^*_{\alpha IK}(\beta,\gamma)\Phi_{\alpha' IK}(\beta,\gamma)=\delta_{\alpha\alpha'},
\end{multline}
where the volume element $|G(\beta,\gamma)|^{1/2} $ is given by
\begin{equation}\label{eq:volume_element}
|G(\beta,\gamma)|^{1/2}=2\beta^4\sqrt{W(\beta,\gamma)R(\beta,\gamma)}\sin 3\gamma,
\end{equation}
with $W(\beta,\gamma)=\{D_{\beta\beta}(\beta,\gamma)D_{\gamma\gamma}(\beta,\gamma) -[D_{\beta\gamma}(\beta,\gamma)]^2\}\beta^{-2}$ and $R(\beta,\gamma)=D_1(\beta,\gamma)D_2(\beta,\gamma)D_3(\beta,\gamma)$.
Using the collective wave functions~\eqref{eq:collective_wavefunction}, 
the reduced quadrupole transition probability is given by
\begin{equation}\label{eq:BE2}
    B(E2; \alpha I \to \alpha^\prime I^\prime ) = (2I+1)^{-1}|\braket{\alpha I||\hat{\mathcal{M}}(E2)||\alpha^\prime I^\prime}|^2,  
\end{equation}
where $\hat{\mathcal{M}}(E2)$ is the electric quadrupole operator.

The collective potential $V(\beta,\gamma)$ is obtained by solving the constrained Hartree--Fock--Bogoliubov (CHFB) equation constructed from a Skyrme energy density functional (EDF).
The vibrational masses and the rotational moments of inertia are calculated by the so-called cranking approximation~\cite{inglis56,beliaev61,girod79} 
with the quasiparticle states obtained by the CHFB equation, hereafter denoted as the cranking mass.

We solve the CHFB equation
with the two-basis method~\cite{gall94,terasaki95} in the three-dimensional Cartesian mesh of $R=12.4$\,fm with the mesh size of $0.8$\,fm. 
The single-particle states at positive energies are cut off so as to become the equivalent quasiparticle energy of $E_{\text{QP}} \approx 30$\,MeV, which gives a good convergence in both the collective potential and the cranking mass in this paper.
We used the SkM$^*$ EDF~\cite{bartel82} and the pairing EDF proposed in Ref.~\cite{yamagami09}, which depends on the isoscalar and isovector densities.
Note that, since we used a different pairing cutoff scheme~\cite{bonche85,ev8} from the one in Ref.~\cite{yamagami09}, we refitted the pairing strength to reproduce the neutron pairing gap of $^{156}$Dy.
For the collective Hamiltonian in the $\beta$--$\gamma$ plane,
we employ a triangular mesh of $\Delta \beta \approx 0.035$ in the region $0<\beta<0.6$ and $0^\circ <\gamma < 60^\circ$, 
consisting of about 200 mesh points.

\section{Results and discussion}\label{sec:result}

\begin{figure}
    \includegraphics[width=\linewidth,clip]{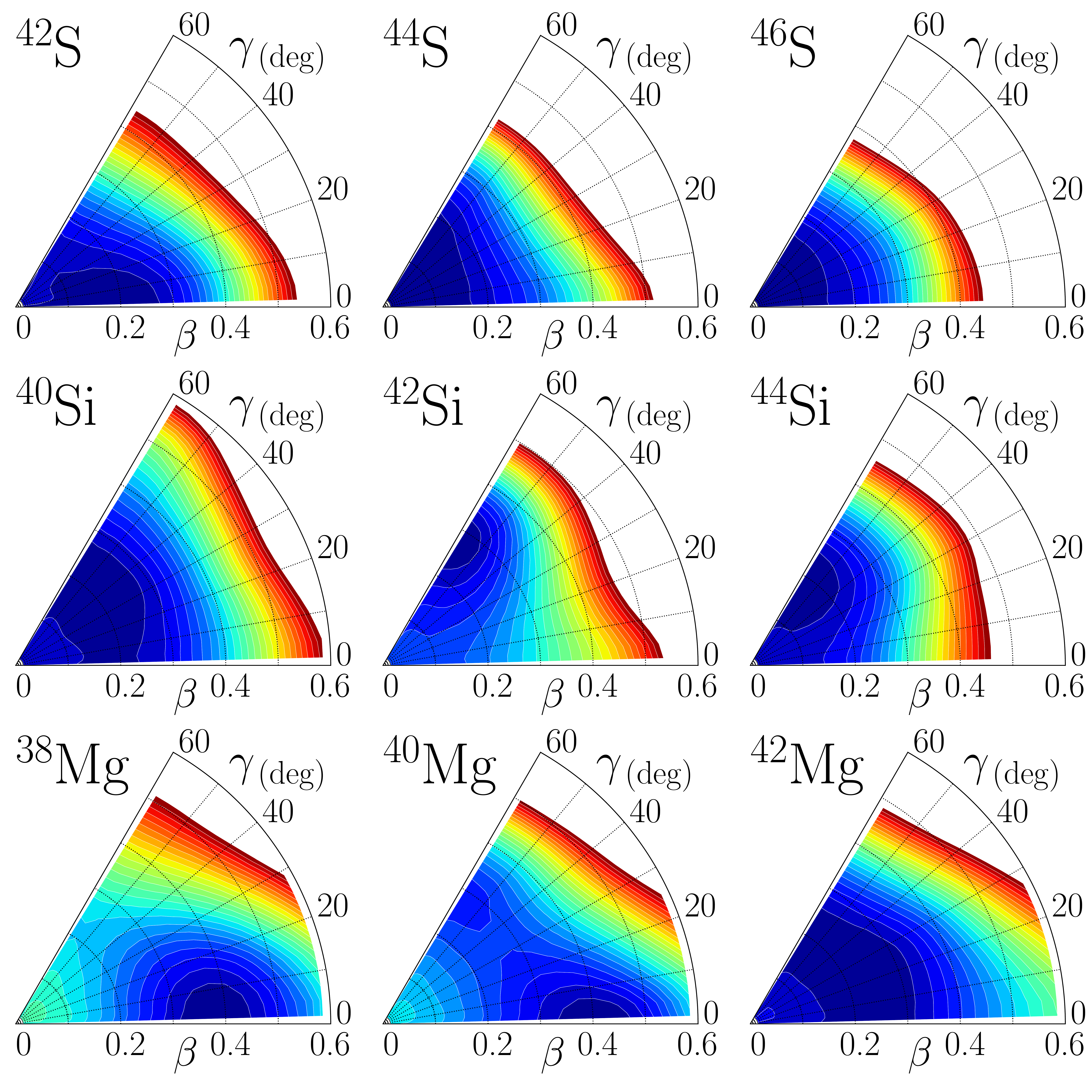}
    \includegraphics[width=0.45\linewidth,clip]{{colorbar}.pdf}
    \caption{
    Potential energy surface in the $\beta$--$\gamma$ plane
    of Mg, Si, and S isotopes with $N=26$, 28, and 30 obtained from
    constrained DFT calculations with the SkM$^*$ EDF.
    }\label{fig:PES}
\end{figure}

\begin{figure}
    \includegraphics[width=0.9\linewidth,clip]{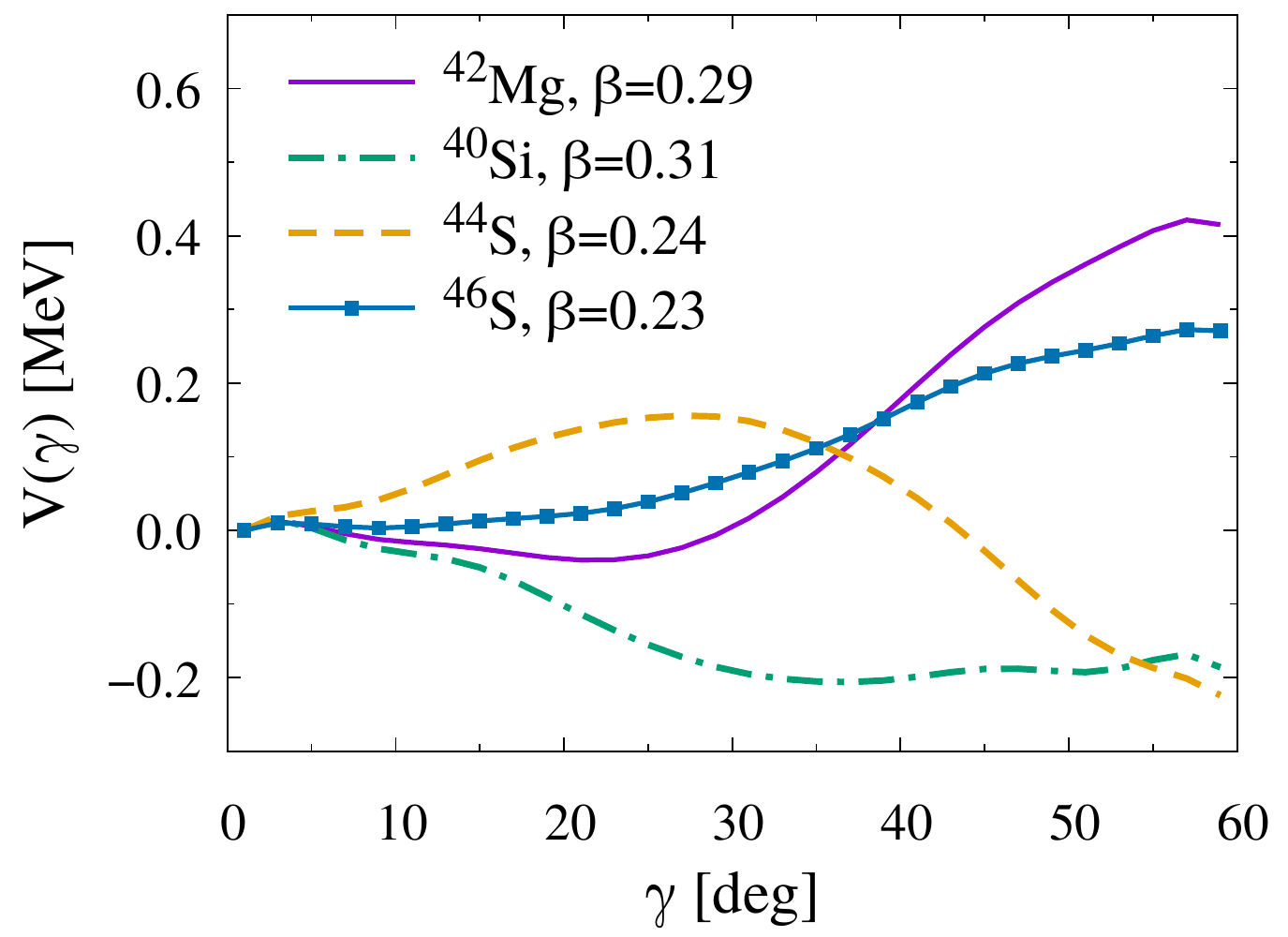}
    \caption{
    Potential energy as a function of $\gamma$ at a fixed $\beta$
    for the selected nuclei. The potential is shifted to $V(\gamma=0^\circ)=0$.
    }\label{fig:pot_gamma}
\end{figure}     

Figure \ref{fig:PES} shows the PESs 
in the $\beta$--$\gamma$ plane 
for the Mg, Si, and S isotopes with $N=26$, 28, and 30 calculated by the constrained HFB. 
In the obtained PESs,
one sees the minimum 
at the prolate configuration in 
$^{38}$Mg and $^{42}$S and 
at the oblate configuration in $^{42,44}$Si.
The PES of $^{40}$Mg has two local minima at the prolate and oblate sides.
One can see that 
$^{42}$Mg, $^{40}$Si, $^{44}$S, and $^{46}$S possess a unique property that
the potential shows a flat behavior from the spherical point to a certain value of $\beta$, 
and along the $\gamma$ direction, namely $\gamma$ soft.

To see clearly a $\gamma$-soft behavior of PESs,
Fig.~\ref{fig:pot_gamma} shows the potential as a function
of $\gamma$ at a fixed $\beta$ for the nuclei mentioned above. 
In this figure, the potential is shifted to become $V(\gamma=0^\circ)=0$.
For each nucleus, the value of $\beta$ is chosen as 
its mean value obtained with the ground-state
collective wave function ($\alpha=1$, $I=K=0$), 
\begin{equation}
    \braket{\beta} = \int d\beta d\gamma |G(\beta,\gamma)|^{1/2} |\Phi_{100}(\beta,\gamma)|^2 \beta,
\end{equation}
and indicated in the figure.
The change in potential along the $\gamma$ direction is 0.2--0.4 MeV.

\begin{figure}
    \includegraphics[width=\linewidth,clip]{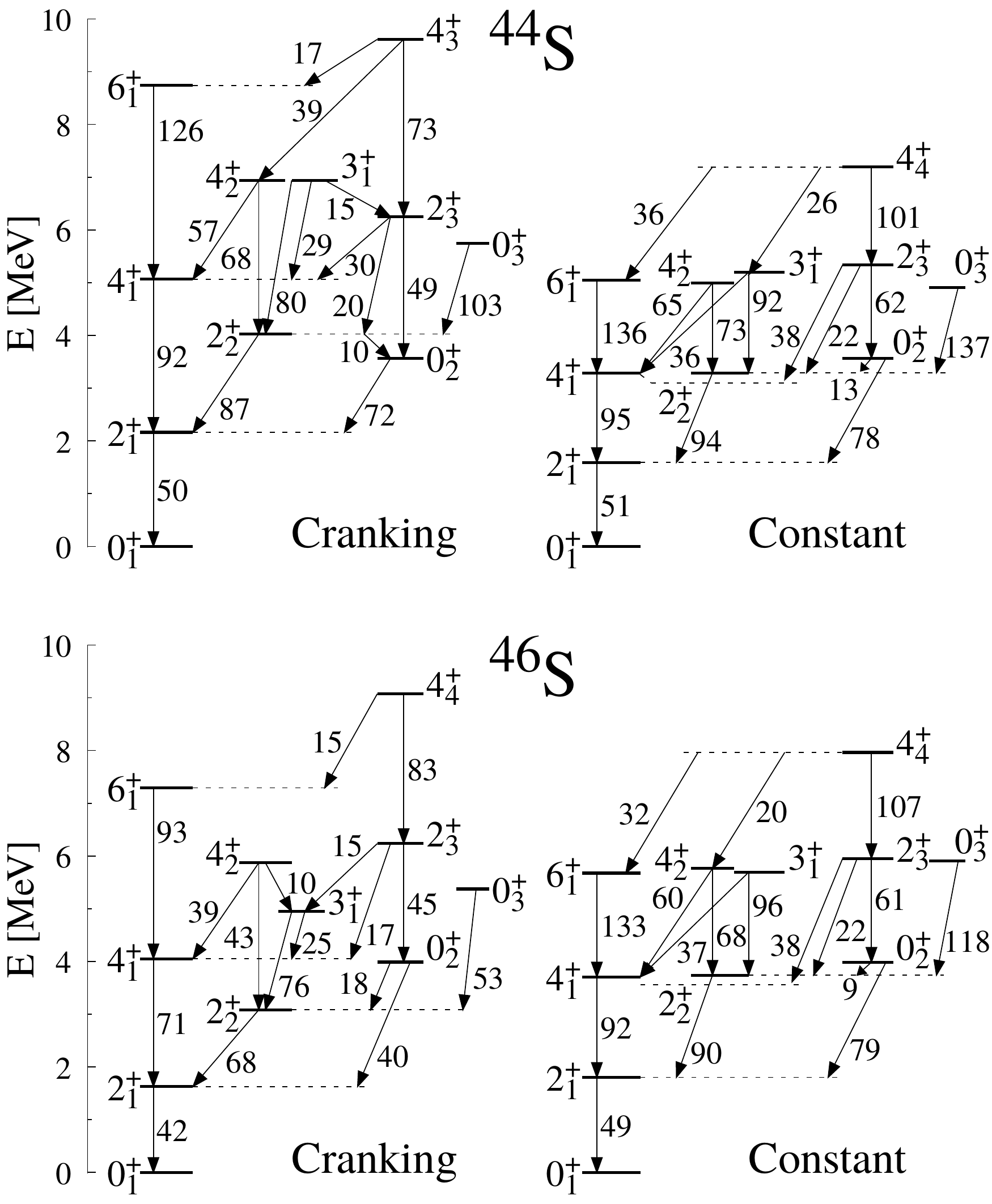}
     \includegraphics[width=\linewidth,clip]{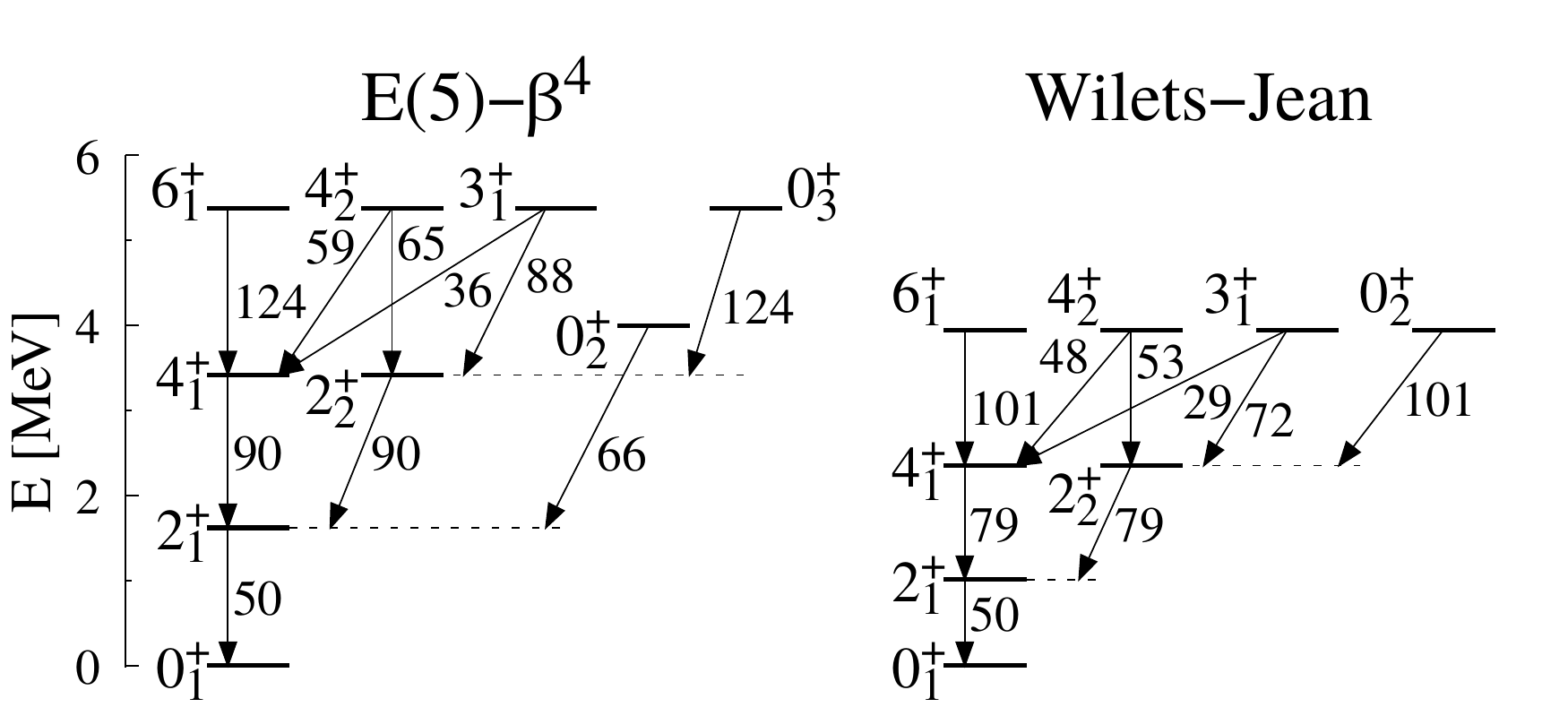}
    \caption{
    Low-lying excitation spectra and $B(E2)$ values in units of $e^2$\,fm$^4$ of $^{44}$S (top) and $^{46}$S (middle) with the cranking mass (left) and the constant mass (right).
    Those of the E(5)--$\beta^4$ and Wilets--Jean models are shown at the bottom. In the E(5)--$\beta^4$ and Wilets--Jean models, the value of $D$ in the constant mass is determined by fitting $E(0_2^+)=4$\,MeV and the $B(E2)$ values are normalized to $B(E2; 2_1^+ \to 0_1^+)=50$\,$e^2$\,fm$^4$.
    }\label{fig:level_BE2}
\end{figure}     

In what follows, we investigate the low-lying states unique 
in $\gamma$-soft nuclei.
The left panels of Fig~\ref{fig:level_BE2} shows low-lying spectra 
of $^{44}$S (top) and $^{46}$S (middle) obtained with the cranking mass.
In the spectra, $I\le 6$ in the $0_1^+$ band and $I\le 4$ in the $0_2^+$ and $2_2^+$ bands are plotted and the $B(E2)$ values larger than 1~W.u. are shown.
We find that the spectra between $^{44}$S and $^{46}$S are very different,
especially the order of $4_1^+$, $2_2^+$, and $0_2^+$,
even though their PESs are very similar to each other as shown in Fig.~\ref{fig:PES}.
The energy ratios defined as $R_{0/2}=E(0_2^+)/E(2_1^+)$, $R_{2/2}=E(2_2^+)/E(2_1^+)$, and $R_{4/2}=E(4_1^+)/E(2_1^+)$ are 
$R_{0/2}=1.65$ and 2.45, $R_{2/2}=1.86$ and 1.89, and $R_{4/2}=2.34$ and 2.49
for $^{44}$S and $^{46}$S, respectively. 
To understand the reason for this difference,
we focus on the role of the vibrational and rotational masses on excitation spectra.
To this end, we employ the Bohr Hamiltonian with a constant mass~\cite{bohr52}, 
where the vibrational and rotational masses obtained by the cranking approximation is
replaced by a constant value, neglecting the $\beta$--$\gamma$ dependence.
Namely~\cite{bohr52},
\begin{equation}
D_{\beta\beta} = D_{\gamma\gamma}/\beta^2=D_1=D_2=D_3 \equiv D, \quad D_{\beta\gamma} = 0.
\end{equation}
The right panels in Fig.~\ref{fig:level_BE2} show the spectra of $^{44}$S and $^{46}$S obtained with the constant mass.
The value of $D$ is determined to reproduce the $0_2^+$ energy obtained with the cranking mass.
Clearly, the patterns of the spectra obtained with the constant mass 
are similar to each other.
The energy ratios are $R_{0/2}=2.24$ and 2.24, $R_{2/2}=2.07$ and 2.08, and $R_{4/2}=2.06$ and 2.06
for $^{44}$S and $^{46}$S, respectively. 
Note that how to determine the value of $D$
does not change these energy ratios, though the absolute value of energy changes. 
If $D$ is determined to give the $2_2^+$ energy with the cranking mass, we obtain 
$R_{0/2}=2.23$, $R_{2/2}=2.07$, and $R_{4/2}=2.06$ for $^{44}$S.

\begin{figure}
\includegraphics[width=\linewidth,clip]{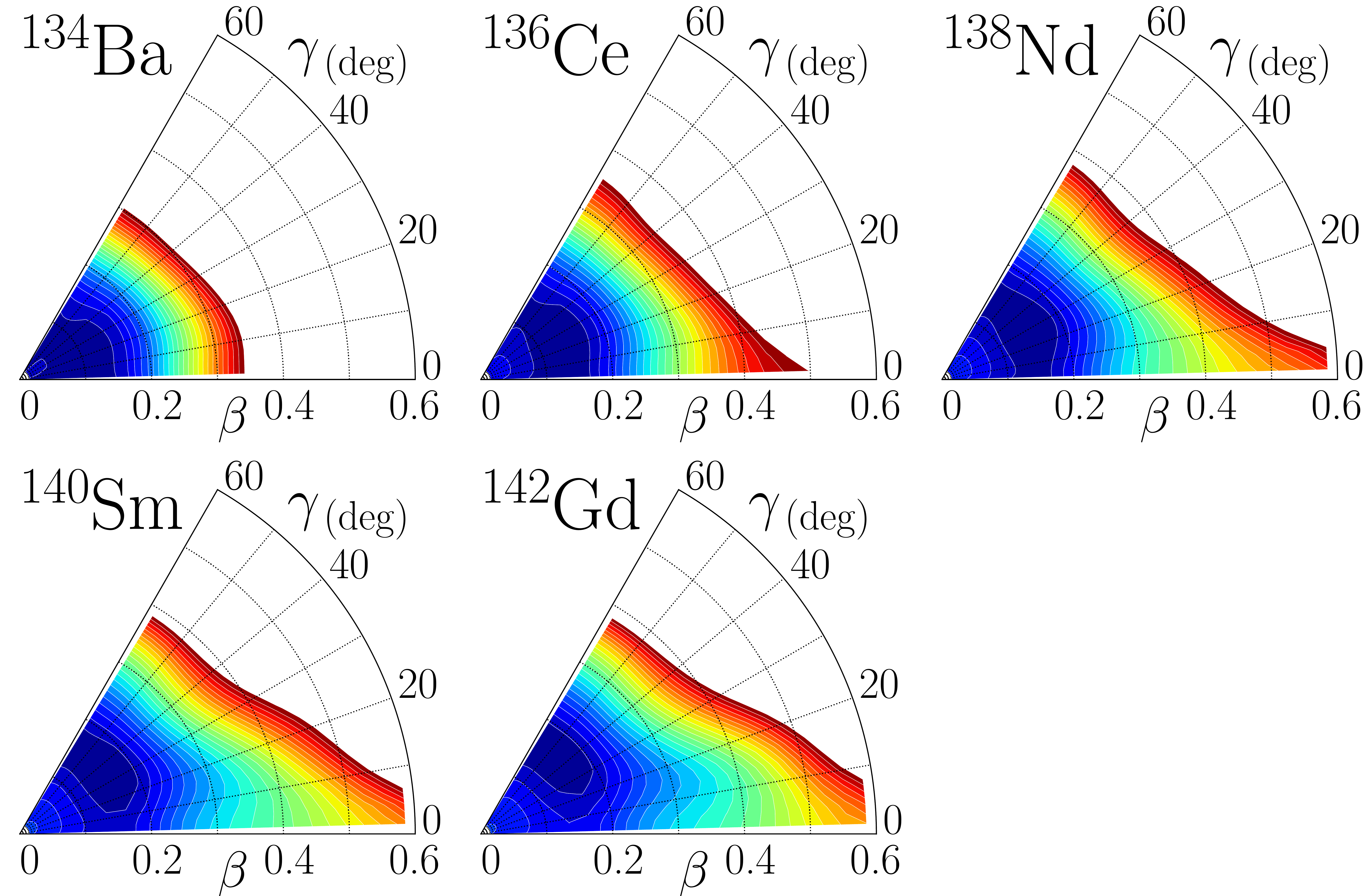} \\
\includegraphics[width=0.45\linewidth,clip]{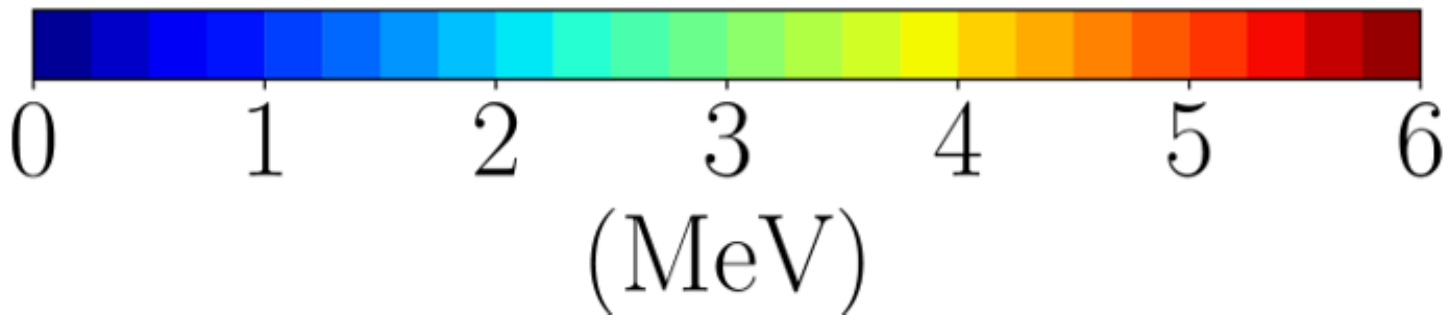} 
\caption{
Same as Fig.~\ref{fig:PES}, but for the neutron-deficient $N=78$ $^{134}$Ba, $^{136}$Ce, $^{138}$Nd, $^{140}$Sm, and $^{142}$Gd nuclides.
}\label{fig:PESN78}
\end{figure}     

\begin{figure}
    \includegraphics[width=0.9\linewidth,clip]{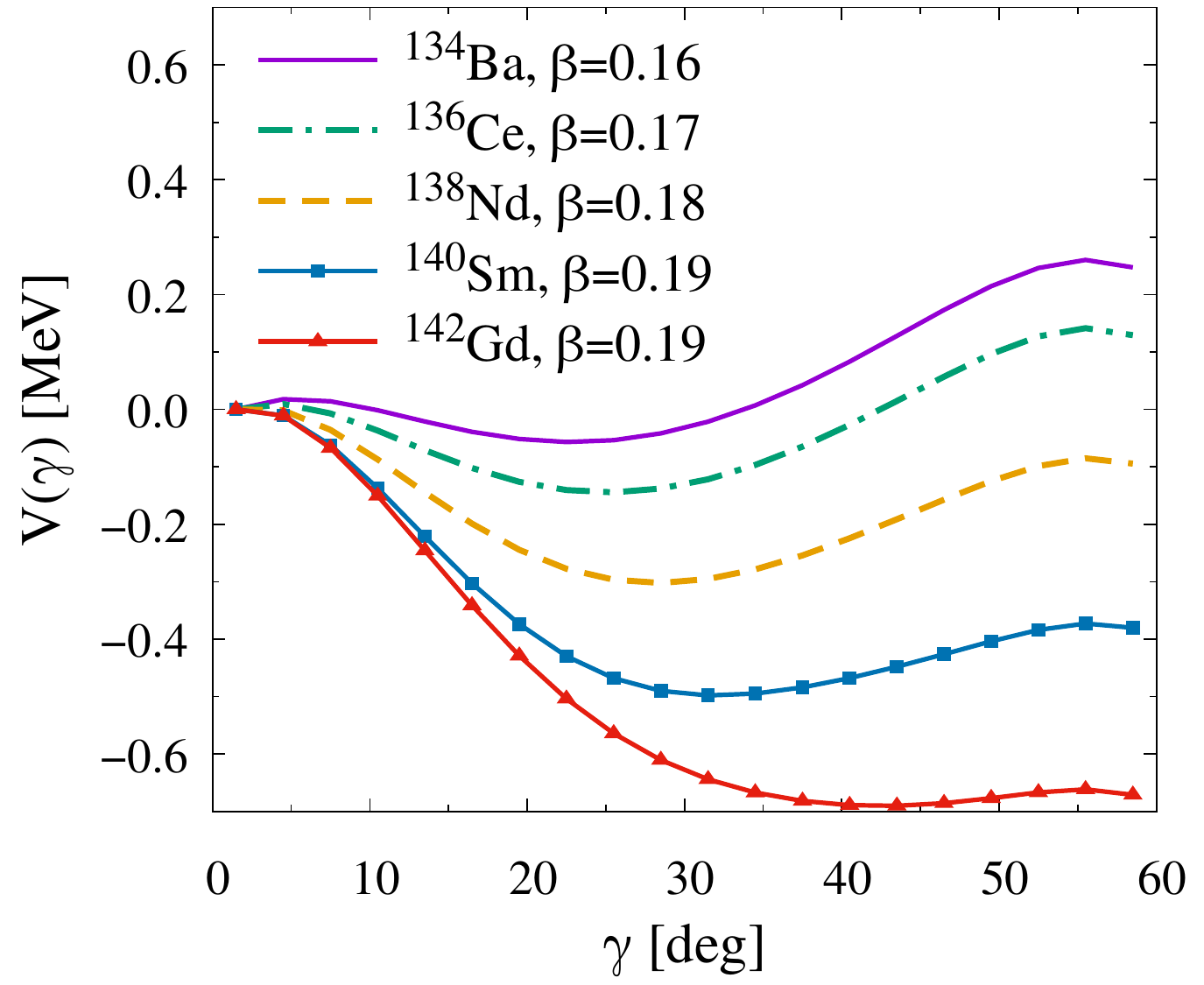}
    \caption{
    Same as Fig.~\ref{fig:pot_gamma},
    but for the $N=78$ nuclei.
    }\label{fig:pot_gamma_N78}
\end{figure}   

To see whether the low-lying spectra are sensitive to the collective mass in other mass regions, 
we study low-energy dynamics of neutron-deficient $N=78$ nuclei $^{134}$Ba, $^{136}$Ce, $^{138}$Nd, $^{140}$Sm, and $^{142}$Gd,
whose PESs show flat in the $\beta$ direction and $\gamma$ soft similarly to those of $^{44}$S and $^{46}$S.
Figures~\ref{fig:PESN78} and \ref{fig:pot_gamma_N78} show
how PESs of those nuclei are $\gamma$ soft. 
The change in PESs is 0.2--0.6 MeV. 
As the proton number decreases, the PES becomes flatter.

\begin{figure}
    \includegraphics[width=\linewidth,clip]{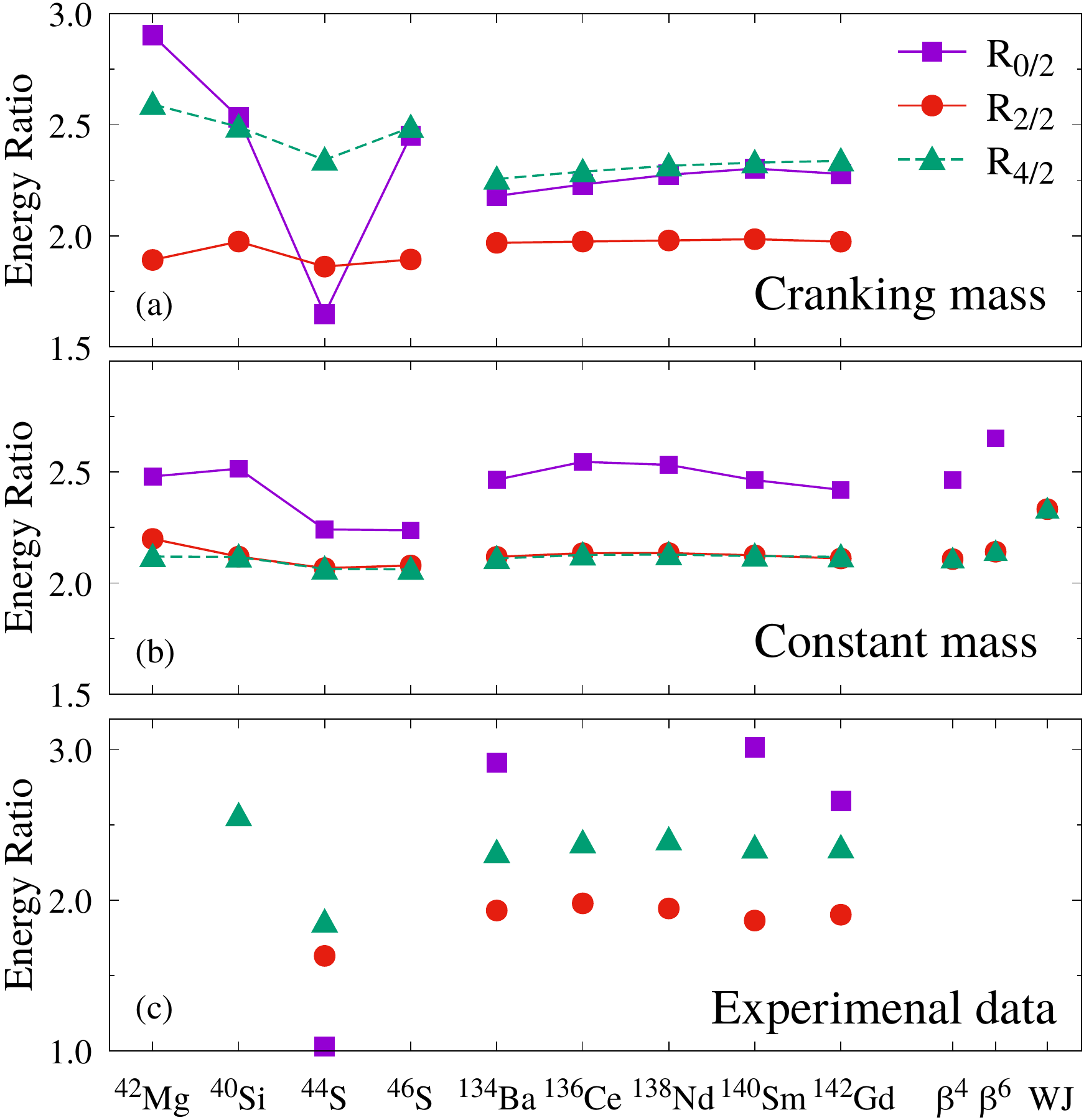}
    \caption{
    $R_{0/2}$ (filled square), $R_{2/2}$ (filled circle) and $R_{4/2}$ (filled triangle) for the selected nuclei 
    with the cranking mass (a) and constant mass (b). The ratios obtained with the E(5)--$\beta^4$
    and E(5)--$\beta^6$ models and the Wilets--Jean model (WJ) are included in (b).
    Note that the $R_{0/2}$ value for the Wilets--Jean model is 3.9.
    The panel (c) shows the ratios of the available experimental data from \cite{santiago-gonzalez11} for $^{44}$S, from \cite{takeuchi12} for $^{40}$Si, and from \cite{NNDC} for the $N=78$ nuclei. 
    }\label{fig:energy_ratio}
\end{figure}     

Figure~\ref{fig:energy_ratio} shows the $R_{4/2}$, $R_{2/2}$, and $R_{0/2}$ values of $^{42}$Mg, $^{40}$Si, $^{44}$S, and $^{46}$S and those of the selected $N=78$ nuclei.
The top panel shows the ratios obtained with the cranking mass, while the bottom depicts those with the constant mass.
In the light nuclei with the cranking mass, $R_{0/2}$ strongly depends on the nucleus, 
while the variation of $R_{4/2}$ and $R_{2/2}$ is small. 
The ratios in $N=78$ nuclei are almost constant, 
$R_{2/2}\sim 2.0$ and $R_{0/2}\sim R_{4/2} \sim 2.2$--2.3 by changing nucleus.
Compared with the cranking mass case,
the variation of $R_{0/2}$ in the light nuclei becomes significantly small in the constant mass case.  
In the $N=78$ nuclei,
$R_{4/2}$, $R_{2/2}$, and $R_{0/2}$ values are almost constant around 2.1, 2.1, and 2.5, respectively.

Properties of the low-energy states in $\gamma$-soft nuclei have often been discussed 
in view of the E(5) critical point symmetry~\cite{iachello00,casten00,casten06}.
The E(5) symmetry is realized in the Bohr Hamiltonian with an infinite 
square-well potential in $\beta$ and a constant in $\gamma$. 
The $\beta^{2n}$ form of the potential instead of the infinite-well potential is introduced to describe realistic systems~\cite{bonatsos04,casten06}.
In Fig.~\ref{fig:energy_ratio}, 
the energy ratios in the E(5)--$\beta^4$ and E(5)--$\beta^6$ models are included and
denoted as $\beta^4$ and $\beta^6$, respectively.
The energy ratios in the light $N\sim 28$ nuclei and $N=78$ nuclei with the constant mass are close to those in the E(5)--$\beta^4$ and E(5)--$\beta^6$ models.
The Wilets--Jean model, which describes $\gamma$-flat PES, gives higher energy ratios than the ones with the constant mass and with the E(5)--$\beta^4$ and E(5)--$\beta^6$ models.
In the bottom panels of Fig.~\ref{fig:level_BE2}, 
the low-lying spectra of the E(5)--$\beta^4$ and the Wilets--Jean models are shown up to seniority three.
These spectra are obtained by $V(\beta,\gamma)=C\beta^4$ with $C=80$\,MeV for E(5)--$\beta^4$ and by $V(\beta,\gamma)=C(\beta-0.3)^2$ with $C=200$\, MeV for the Wilets--Jean model,
and with the constant mass $D$ giving $E(0_2^+)=4$\,MeV.
The pattern of low-lying spectra in the E(5)--$\beta^4$ is close to those of $^{44}$S and $^{46}$S with the constant mass: 
The degeneracy of the $2_2^+$ and $4_1^+$ states and that of the $0_3^+$, $3_1^+$, $4_2^+$, and $6_1^+$ states.

The energy ratios obtained from the available experimental data in neutron-rich $N\sim 28$ nuclei are 
$R_{4/2}=1.86$, $R_{2/2}=1.63$ and $R_{0/2}=1.03$ in $^{44}$S~\cite{santiago-gonzalez11}, and $R_{4/2}=2.56$ in $^{40}$Si~\cite{takeuchi12}, 
as shown in Fig.~\ref{fig:energy_ratio}(c).
The energy ratio $R_{0/2}$ with the cranking mass shows a strong nucleus dependence in neutron-rich $N\sim 28$ nuclei, and 
one sees a sudden drop in $^{44}$S although 
the measured value is even lower than 
the calculated one. 
In the $N=78$ nuclei, the energy ratios of the available experimental data in Fig.~\ref{fig:energy_ratio}(c) are almost constant, though some of the $R_{0/2}$ values are not available.
These energy ratios are similar to those with both the cranking and constant mass cases in $N=78$.
More experimental data on the low-lying $0^+$ state 
will give insight into roles of the collective mass.

\begin{figure}
    \includegraphics[width=0.9\linewidth,clip]{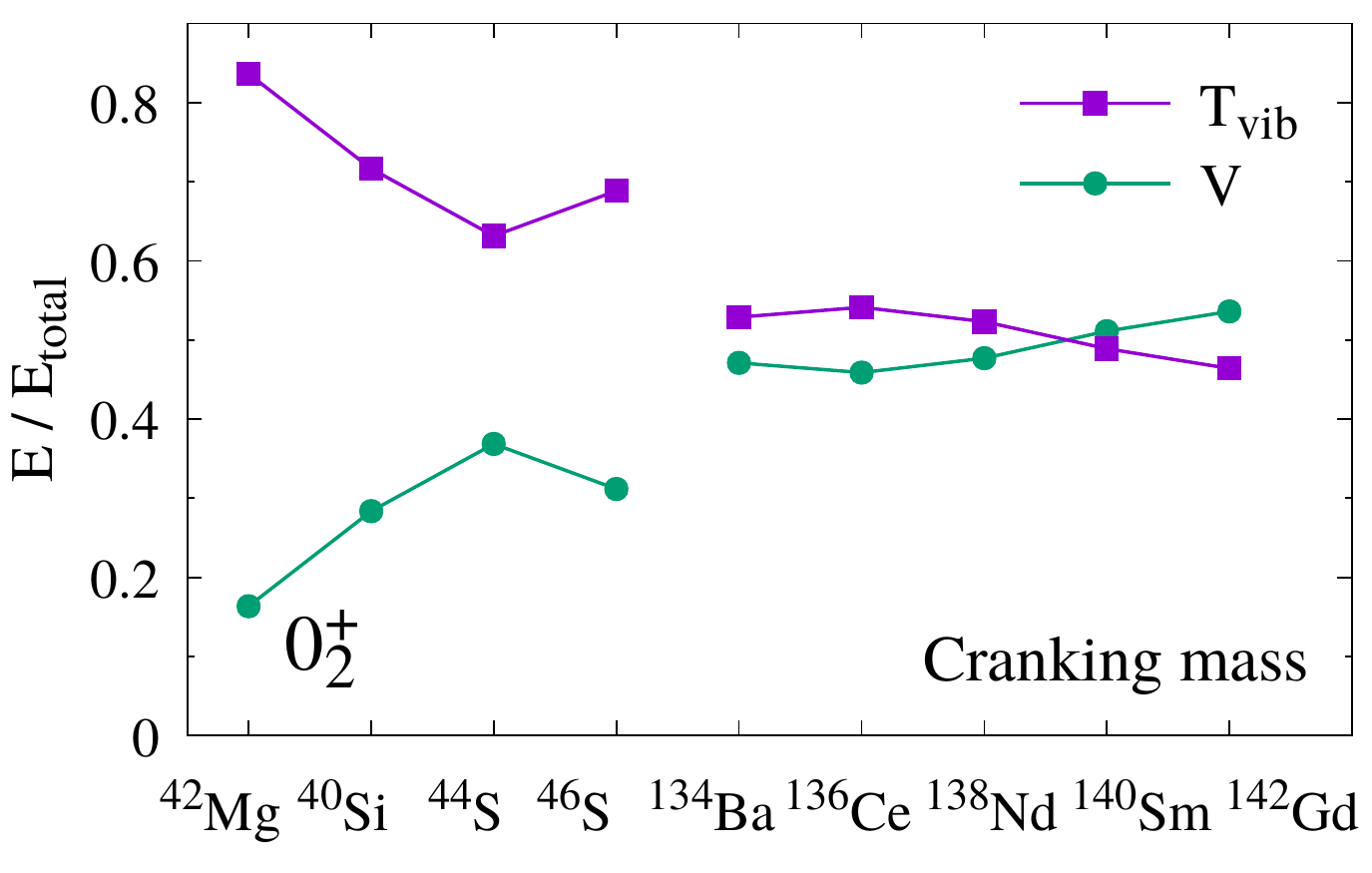}
    \caption{
    Decomposition of the $0_2^+$ energy to the vibrational kinetic energy $T_{\text{vib}}$ and the potential energy $V$ divided by the total energy with the cranking mass for the selected nuclei.
    }\label{fig:energy_expectation_decomposition}
\end{figure}    

To further investigate the origin of the different behaviors of $R_{0/2}$ in the $N \sim 28$ and $N=78$ nuclei,
we decompose the energy of the $0_2^+$ state to the vibrational kinetic energy and potential energy. 
These energies are calculated 
with the vibrational wave function of the $0_2^+$ state as 
\begin{equation}\label{eq:energy_expectation_decomposition}
    E = \int d\beta d\gamma |G(\beta,\gamma)|^{1/2} \Phi^*_{200}(\beta,\gamma) \hat{E} \Phi_{200}(\beta,\gamma),  
\end{equation}
where $\hat{E} = \hat{T}_{\text{vib}}$ or $V(\beta,\gamma)$.
Figure~\ref{fig:energy_expectation_decomposition} shows the vibrational kinetic energy $T_{\text{vib}}$ and the potential energy $V$ divided by their sum $E_{\text{total}} = \braket{T_{\text{vib}}} + \braket{V}$, the absolute value of the $0_2^+$ energy,
for the $N\sim 28$ and $N=78$ nuclei. 
Note that there is no contribution of rotation 
in the $I=0$ states.
In the $N\sim 28$ nuclei, $T_{\text{vib}}$ is larger than $V$. 
A large $T_{\text{vib}}$ with the deformation-dependent cranking mass 
enhances the nucleus dependence of the $0_2^+$ energy.
Furthermore, we found that 
the ratios of $T_{\text{vib}}$, $T_{\text{rot}}$, and 
$V$ to the total energy 
in the $0_1^+$ and $2_1^+$ states
are almost the same and do not depend on 
the nucleus so much.
Therefore, we observe the
strong nucleus-dependent $R_{0/2}$ values. 
In $N=78$, the two energies are close to each other and do not much depend on the nucleus.
Thus, we have obtained
that the $R_{0/2} $ values do not depend on the nucleus
even with the
deformation-dependent cranking mass.

\begin{figure}
    \includegraphics[width=\linewidth,clip]{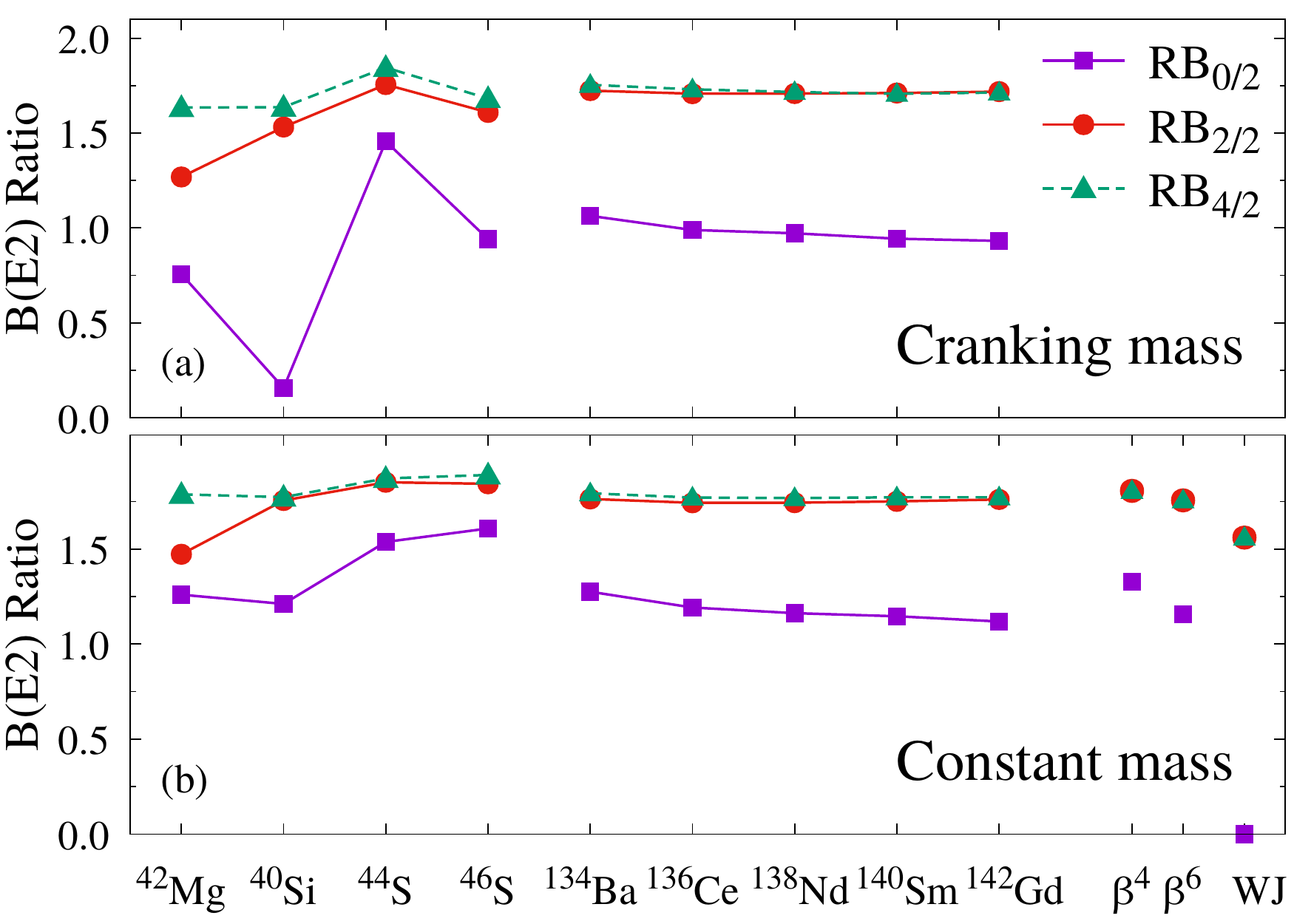}
    \caption{ 
    $B(E2)$ ratio for the selected nuclei with the cranking mass (a)
    and constant mass (b) together with E(5)--$\beta^4$ and E(5)--$\beta^6$ models and the Wilets--Jean model (WJ). 
    }\label{fig:BE2_ratio}
\end{figure}    

\begin{figure*}
    \includegraphics[width=0.96\linewidth,clip]{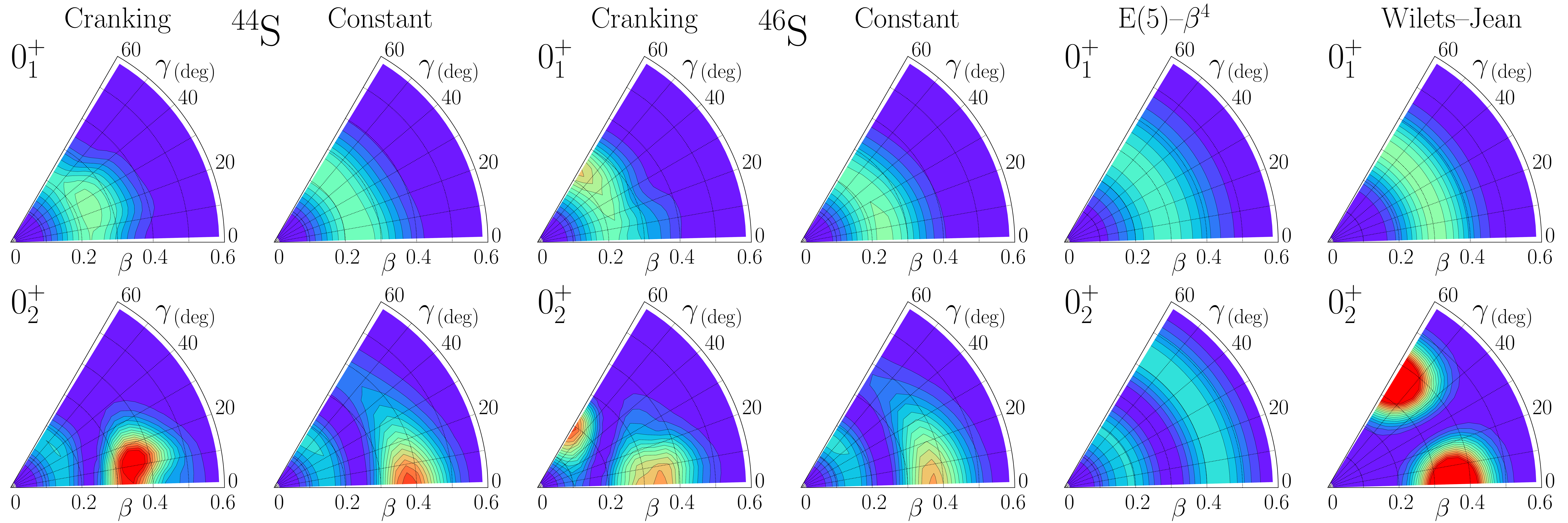}  
    \includegraphics[width=0.02\linewidth,clip]{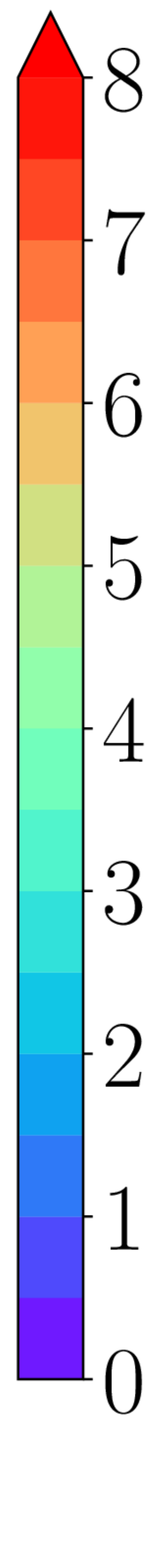}
    \caption{ 
    Vibrational wavefunctions $|\Psi_{\alpha 00}(\beta,\gamma)|^2$ multiplied by $\beta^4 |W(\beta,\gamma)R(\beta,\gamma)|^{1/2}$
    of the $0_1^+$ (top) and $0_2^+$ (bottom) states with the cranking and constant masses for $^{44}$S and $^{46}$S.
    The right two columns show those with the E(5)--$\beta^4$ and Wilets--Jean models.
    }\label{fig:wavefunction}
\end{figure*}

The similar feature to the energy ratios is indeed obtained in the $B(E2)$ ratios.
Figure~\ref{fig:BE2_ratio} shows the $B(E2)$ ratios
$RB_{0/2}=B(E2;0_2\to 2_1)/B(E2;2_1\to 0_1)$,
$RB_{2/2}=B(E2;2_2\to 2_1)/B(E2;2_1\to 0_1)$, and 
$RB_{4/2}=B(E2;4_1\to 2_1)/B(E2;2_1\to 0_1)$
for the selected nuclei 
with the cranking mass (top) and the constant mass (bottom).
The $RB_{0/2}$ value in the light $N\sim 28$ nuclei with the cranking mass
shows a strong dependence on the nucleus.
The change in the $RB_{0/2}$ value in the light nuclei with the constant mass is relatively moderate.
In the $N=78$ nuclei, however, 
all the $B(E2)$ ratios considered do not depend on the nucleus with both the cranking mass and the constant mass.
The $B(E2)$ ratios in both the light $N\sim 28$ nuclei and $N=78$ nuclei with the constant mass are close to those in the E(5)--$\beta^4$ and E(5)--$\beta^6$ models.
These results imply that the $0_2^+$ state in the light nuclei is sensitive to the collective mass. We are going to investigate the $0_2^+$ state in terms of the collective wave functions below.
Furthermore, we conclude that the sensitivity of the collective mass to the low-lying spectra is strong in the light neutron-rich nuclei around $N=28$ uniquely.

We discuss the structure of the collective wave functions (WFs) of the $0_1^+$ and $0_2^+$ states.
First, we look into the WFs with the constant mass shown in Fig.~\ref{fig:wavefunction}. 
Here the WFs are multiplied by $\beta^4 \sqrt{W(\beta,\gamma)R(\beta,\gamma)}$, the volume element without $\sin3\gamma$. 
The WF of the $0_1^+$ state 
is spread over along the $\gamma$ direction. This is indeed expected from the $\gamma$-soft property in the PESs. 
The WF of $0_2^+$ has a node along the $\beta$ direction.
This is common in $^{44}$S and $^{46}$S.
As expected from the PESs in $^{44}$S and $^{46}$S, the WFs with the constant mass look similar to those in the E(5)--$\beta^4$ model shown in the
fifth column in Fig.~\ref{fig:wavefunction}.
Next, we discuss the WFs with the cranking mass.
The WF of the $0_1^+$ state is spread along the $\gamma$ direction 
as obtained with the constant mass and peaks at $\gamma\sim 20^\circ$ in $^{44}$S and at the oblate side in $^{46}$S.
The WF of the $0^+_2$ state in $^{44}$S is more or less similar to that obtained with the constant mass,
although the localization around the prolate side is strong.
In $^{46}$S, the structure of the WF of $0_2^+$ is different from that obtained with the constant mass: 
The WF has two peaks at the prolate and oblate sides. 
The two-peak structure in the collective WF of the $0^+_1$ and $0^+_2$ states 
is a typical feature of shape coexistence, such as in $^{72}$Kr~\cite{sato11}.
The collective WF of the only $0^+_2$ state in $^{46}$S looks similar to 
the collective WFs of the $0^+_1$ and $0^+_2$ states in $^{72}$Kr. 
Thus, the $0^+_1$ and $0^+_2$ states in $^{46}$S are not interpreted 
as a usual shape coexistence.
This result is rather similar to the one obtained in the Wilets--Jean model 
shown in the sixth column in Fig.~\ref{fig:wavefunction}.

\section{Summary}\label{sec:summary}
We have investigated the role of the mass parameters in the collective Hamiltonian for the triaxial-shape dynamics in neutron-rich nuclei with $N\simeq 28$.
The PESs are obtained by the constrained HFB method 
with a Skyrme-type EDF.
We found that the PES in $^{42}$Mg, $^{40}$Si, $^{44}$S, and $^{46}$S 
possesses a topography similar to each other and 
is soft against triaxial deformation.
The low-lying spectra obtained by assuming the mass parameters as constant are similar.
However, the spectra obtained considering the deformation dependence 
of the mass parameters with the cranking approximation show characteristic features. 
The second $0^+$ state is sensitive to the treatment of the mass parameters.
The energy ratio $R_{0/2}$ and the $B(E2)$ ratio $RB_{0/2}$ show a strong nucleus dependence.
The dependence of $R_{0/2}$ and $RB_{0/2}$ on the nucleus in neutron-deficient $N=78$ nuclei, which also exhibit the $\gamma$-soft nature,  
is less pronounced.
We clarified the unique
role of the collective mass in $N\sim 28$ nuclei.

\section*{Acknowledgments}
The authors thank M.~Kimura and Y.~Suzuki for the valuable discussions. 
This work was supported by 
JSPS KAKENHI (Grants No. JP19K03824 and No. JP19K03872) 
and the JSPS/NRF/NSFC A3 Foresight Program ``Nuclear Physics in the 21st Century.''

%

\end{document}